\documentclass[%
 footinbib,
 twocolumn,
 superscriptaddress,
 longbibliography,
 amsmath,amssymb,
 aps,
 floatfix, 
 prb,
 floatfix,
 color,
 nobalancelastpage,
]{revtex4-1}

\usepackage{physics}
\usepackage[version=4]{mhchem}
\usepackage{color}
\usepackage{graphicx}
\usepackage{hyperref}
\usepackage{cleveref} %last import, clever references; use \cref{}

\renewcommand{\=}{\!=\!}

\crefname{equation}{Eq.}{Eqs.}
\crefname{figure}{Fig.}{Figs.}
\crefname{section}{Sec.}{Secs.}
\Crefname{equation}{Equation}{Equations}
\Crefname{figure}{Figure}{Figures}
\Crefname{section}{Section}{Sections}

\hypersetup{colorlinks,allcolors=black}

\begin{document}

\title{Impact ionization and multiple photon absorptions in the two-dimensional photoexcited Hubbard model}
\author{Florian Maislinger}
\affiliation{Institute of Theoretical and Computational Physics, Graz University of Technology, 8010 Graz, Austria}

\author{Hans Gerd Evertz}
\affiliation{Institute of Theoretical and Computational Physics, Graz University of Technology, 8010 Graz, Austria}

\date{31 July 2020}

\begin{abstract}
  We study the non-equilibrium response of a $4 \times 3$ Hubbard model at $U = 8$ under the influence of a short electric field pulse, with the main focus on multiple photon excitations and on  the change of double occupancy after the pulse.
  The behavior mainly depends on the driving frequency of the electric field.
  The largest change of double occupancy occurs during the pulse.
  For frequencies below the Mott gap, we observe multiphoton excitations at large field intensities.
  For frequencies beyond the gap energy, there is a region where Auger recombination reduces the double occupancy after the pulse.
  Impact ionization (Multi Exciton Generation),  namely a growing double occupancy after the pulse, occurs for frequencies larger than twice the Mott gap.
  From the Loschmidt amplitude we compute the eigenstate spectrum of the quantum state after the pulse, observing multiple distinct photon excitation peaks,
  in line with expectations from a quasiparticle picture.
  We introduce a technique with which we analyze the time evolution of double occupancy in each peak individually.
  The long-term behavior of the double occupancy almost only depends on the absorbed energy, and we explore the connection of this property to the Eigenstate Thermalization Hypothesis.
\end{abstract}

\maketitle

%=================================================
\section{Introduction}
%=================================================

One exciting area of research is the influence of photoexcitations on strongly correlated electron systems.
In strongly correlated materials, one generally cannot neglect the electron-electron interaction when describing and predicting material properties, because strong localization and Coulomb interaction effects play an important role \cite{bickers1989conserving}.
In the context of computational material science this means that one of its most prominent approaches, the Density Function Theory (DFT)~\cite{jones2015density}, fails to reliably predict the correct properties~\cite{bickers1989conserving}.
There are several possible but computationally expensive routes one can take to tackle this problem. One of them is to map the many body Hamiltonian to a Hubbard type model and to use Dynamical Mean Field Theory (DMFT)~\cite{georges1996dynamical} to solve it. The approach we choose here is to consider small lattice sizes,
akin to nano crystals or quantum dots~\cite{schaller2004high,ellingson2005highly,semonin2011peak,pietryga2016spectroscopic},
and to use precise techniques related to exact diagonalization
 to compute the ground state and the time evolution.

An important effect for, e.g., the solar power industry is impact ionization~\cite{manousakis2010photovoltaic,assmann2013oxide,manousakis2019optimizing}, 
also called Multiple Exciton Generation (MEG)~\cite{manzi2018resonantly,wang2017boosting,pietryga2016spectroscopic,goodwin2018multiple}, 
Carrier Multiplication (CM)~\cite{pietryga2016spectroscopic},
or Multi Carrier Generation (MCG)~\cite{manousakis2019optimizing}, 
where an excited electron with
 kinetic energy larger than the band or Mott gap of the material
generates additional excitations through electron-electron scattering.
Impact ionization has been observed experimentally~\cite{holleman2016evidence,franceschetti2006impact,wang2013tuning,wang2017boosting,goodwin2018multiple,manzi2018resonantly,sahota2019manybody}.
It would allow to raise the theoretical efficiency limit of solar cells from approximately $30\%$~\cite{shockley1961detailed} to approximately $60\%$~\cite{landsberg1993band}.
However, impact ionization cannot be used to significantly increase the efficiency of classical silicon based solar cells~\cite{wolf1998solar}, because the electron-phonon scattering process is faster by an order of magnitude than the relevant electron-electron scattering processes~\cite{kirk2012fundamental}.
In strongly interacting Mott insulators, on the other hand, the relevant electron-electron scattering processes can be much faster than electron-phonon scattering processes due to  strong localization and Coulomb interaction mentioned above.
It has therefore been proposed to use Mott insulators for photovoltaic purposes~\cite{manousakis2010photovoltaic,assmann2013oxide,coulter2014optoelectronic}.

Impact ionization has been numerically confirmed to exist for the infinite dimensional hypercubic lattice in a DMFT calculation~\cite{werner2014role} and a quantum Boltzmann approach~\cite{wais2018quantum}, 
and in DMFT calculations for a correlated layer connected to two metallic leads~\cite{sorantin2018impact} and for models of \ce{LaVO3} and \ce{YTiO3}~\cite{petocchi2019hund}.
The drawback of DMFT is that it neglects spatial correlations~\cite{georges1996dynamical}.
In the present paper, 
we take such correlations into full account.
We examine the $4 \times 3$ Hubbard model, exposed to a short laser pulse~\cite{takahashi2008photoinduced,werner2014role,eckstein2013photoinduced,hashimoto2016photoinduced,eckstein2011thermalization,innerberger2020electron,innerberger2020electron,kauch2020enhancement}.
We use L\'{a}nczos-based methods~\cite{lanczos1950iteration} to compute the ground state, and a related scaling method~\cite{almohy2001computing} to compute the time evolution.
An independent study by Kauch et. al.~\cite{kauch2020enhancement} 
observed impact ionization in Hubbard models, with the  focus on different geometries, disorder, and the quasiparticle spectrum.

When the intensity of the electric field is high, one enters the domain of nonlinear optics, where multiphoton absorption processes ~\cite{he2008multiphoton,zipfel2003nonlinear,helmchen2005deep,sun2006twophoton,li2007multiphoton,zhang2010designable,sohn2019observation,albota1998design,banfi1994two,wang2015nonlinear,xu2016twophoton,ogasawara2000ultrafast,ashida2002dimensionality,manzi2018resonantly}
become relevant.
With multiphoton absorption 
it is possible to reach excited states beyond a gap,
even when the energy of a single photon is smaller than the gap.
Multiphoton absorption is a single physical process 
described by Quantum Electrodynamics~\cite{carusotto1967twophoton}, different from the consecutive absorption of single photons~\cite{he2008multiphoton}.
Experimentally, the necessary field strengths are achieved by focusing a pulsed laser beam onto a very small (order of $10^{-9}~\text{cm}^2$) area~\cite{zipfel2003nonlinear}.
Relevant applications include multiphoton microscopy~\cite{zipfel2003nonlinear,helmchen2005deep} and high-resolution three dimensional polymerization of photoresists~\cite{sun2006twophoton,li2007multiphoton,zhang2010designable}. High resolution in comparison to single photon absorption is achieved because of the nonlinear dependence of multiphoton absorption on the field intensity~\cite{sohn2019observation,zhang2010designable}.
The effect can be utilized in a variety of materials including molecules~\cite{albota1998design,he2008multiphoton}, nano-crystals~\cite{banfi1994two,wang2015nonlinear,xu2016twophoton}, cuprates~\cite{ogasawara2000ultrafast,ashida2002dimensionality}, 
and chalcogenide glasses~\cite{sohn2019observation}.
We observe multiphoton absorption at high electric field intensity with a driving frequency below the Mott gap size.

An interesting phenomenon that has been observed for many quantum systems is that the long time mean of the expectation value of an observable can tend to a value which depends only on the energy of the initial state.
This is the topic of the so called Eigenstate Thermalization Hypothesis (ETH)~\cite{srednicki1994chaos,deutsch2018eigenstate}.
The dependence on energy only can be understood when the initial state is dominated by  a single peak in the eigenstate spectrum of the Hamiltonian and the relevant observable varies slowly in eigenenergy.
In the present paper, we observe and explain a similar dependence on energy only, for states with support in a very large energy range.

In \cref{sec:model} we provide an overview over the model, 
briefly present numerical methods, and discuss expectations based on a quasiparticle picture.
\Cref{sec:time_evolution_double_occupancy} shows the time dependence of the double occupancy during and after the photo pulse, including impact ionization.
We calculate the emerging eigenstate spectrum from the Loschmidt amplitude in \cref{sec:eigenstate_spectrum}, observing a clear peaked structure with distances in multiple of the photon energy. The time evolution of the individual peaks is analyzed in \cref{sec:TimeEvolIndividPeaks}.
The long time behavior of the double occupancy is mostly governed by the amount of absorbed energy during the pulse.
A connection to the Eigenstate Thermalization Hypothesis is explored in \cref{sec:eigenstate_thermalization}.

%=================================================
\section{Model} \label{sec:model}
%=================================================

\begin{figure} %====================================================
    \centering
    \includegraphics[width=8.6cm]{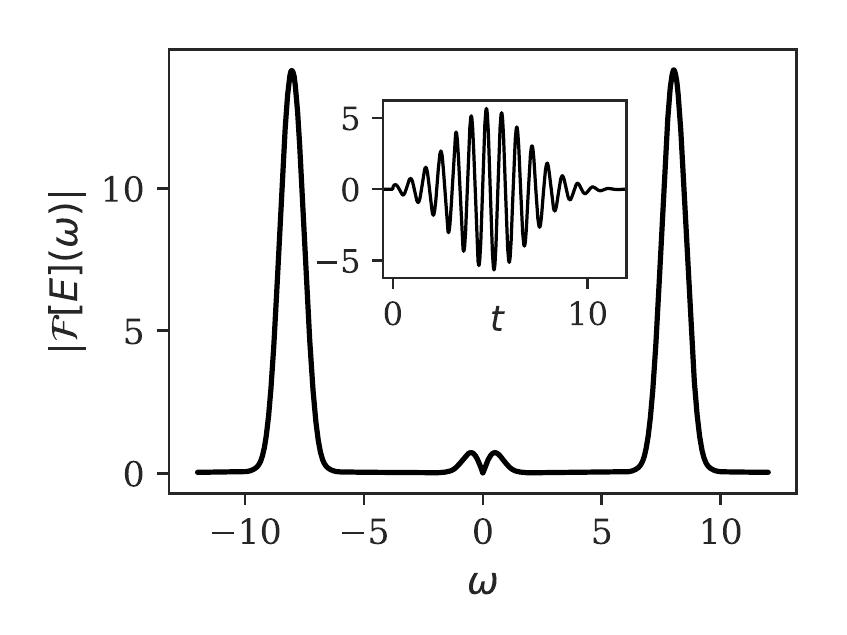}
    \caption{
        Electric field pulse,  for $\Omega = 8$ and $I_0 = 0.71$.
        Inset: time domain. Main figure: 
        absolute value of the Fourier transform of the electric field.
        The two peaks are of Gaussian shape with a width of $\tilde{\sigma} = \frac{1}{2}$.
        \label{fig:electric_field}
    }
\end{figure} %====================================================

\begin{figure} %====================================================
    \centering
    \includegraphics[width=8.6cm]{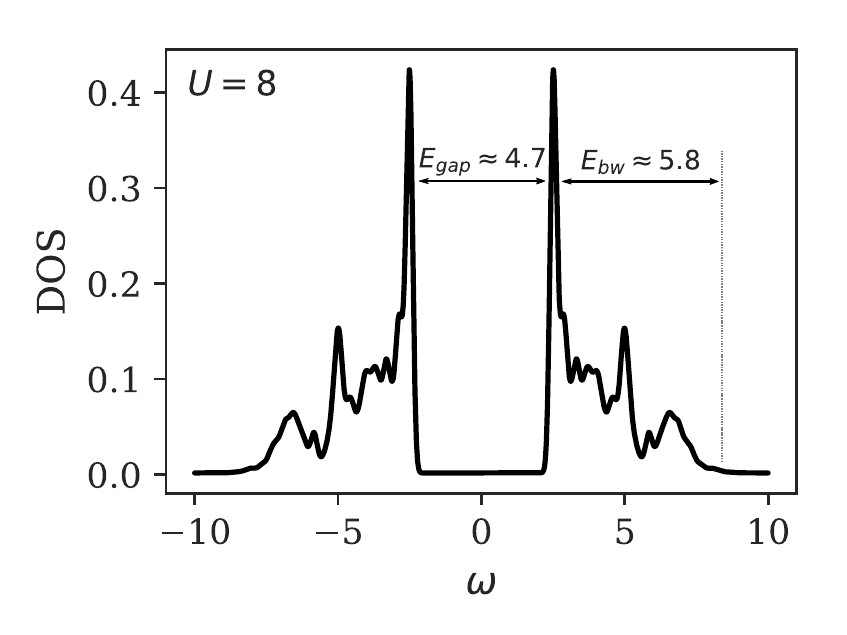}
    \caption{
        Equilibrium density of states of $H(0)$,
        at half filling and $U = 8$ on the $4\times3$ lattice,
        convoluted with a Gaussian peak of width $\sigma = \frac{1}{10}$
        for smoothness.
        \label{fig:hubbard_4x3_dos}
    }
\end{figure} %====================================================

We investigate a $4 \times 3$ Hubbard model % $N=12$ sites
with open boundary conditions and nearest neighbor hopping $v_{ij}(t)$,
\begin{subequations}
    \begin{align}
      H(t) &= H_{0}(t) + H_{1} \\
      H_{0}(t) &= -\sum_{\langle i j\rangle, s} v_{ij}(t) ~ c_{i, s}^{\dagger} c_{j, s} \\
      H_{1} &= U \sum_{i} \left(n_{i,\uparrow} - \frac{1}{2}\right) \left(n_{i,\downarrow} - \frac{1}{2}\right)~
    \end{align}
\end{subequations}
at half-filling, $\sum_i n_{i, \uparrow} = \sum_i n_{i, \downarrow} = \frac{N}{2}$, $N=12$.

The initial state at time $t=0$ is the ground state,
calculated with the L\'{a}nczos method~\cite{lanczos1950iteration}.
At time $t>0$ the system is irradiated by an electric field $\vec E$ of frequency $\Omega$, in plane with the $2d$ system, and angled at $45^{\circ}$ to the lattice.
We incorporate the field into the Hubbard model by the Peierls phase~\cite{peierls1933theorie,wissgott2012dipole},
 $\phi(t) = \int_{x_i}^{x_j} \vec A(x, t) ~ \dd{\vec x}$, 
 where  $\vec A$ is the vector potential, with % gauge choice
  $\vec E = -\frac{\partial \vec A}{\partial t}$.
 Then all nearest neighbor hopping amplitudes obtain the same time-dependent complex phase factor
\begin{equation}
  v_{ij}(t) = v~ e^{i \phi(t)} ~. \\
\end{equation}
We employ a Gaussian shaped light pulse~\cite{werner2014role,hashimoto2016photoinduced,innerberger2020electron,kauch2020enhancement},
centered at time $t_i=5 v^{-1}$, with a width of $\sigma = 2 v^{-1}$
and intensity $I_0$, 
see \cref{fig:electric_field},
so that 
\begin{equation}
  \phi(t)
 = -I_0 ~ e^{-\frac{\left(t - t_i\right)^2}{2 \, \sigma^2}} % &\, 
 \left( \cos\left(\Omega  \left(t - t_i\right) \right) - \cos\left(-\Omega \, t_i \right) \right) ~.
 %    \sigma = 2.0 &t_i = 5
\end{equation}

We investigate the model at $U=8v$.
In the following, all energies will be specified in units of the hopping amplitude $v$ and all times in units of $v^{-1}$.

Our main observable of interest is the total double occupancy of the system:
\begin{equation}
    d = \sum_i \, n_{i, \uparrow} \, n_{i, \downarrow}
\end{equation}
as a measure of excitations of the system.
We will use an increasing double occupancy after the incoming photon pulse as a measure for impact ionization~\cite{werner2014role}.

We approximate the time-evolution operator $\mathcal{U}$ by

\begin{subequations}
    \begin{align}
        \mathcal{U}(t_0 + \Delta t, t_0) &= \mathcal{T} e^{-i \int_{t_0}^{t_0 + \Delta t} H(\bar{t}) d\bar{t}} \\
        &\approx e^{-\frac{i \Delta t}{2} \left(H(t_0) + H(t_0 + \Delta t)\right)}
    \end{align}
\end{subequations}

We use a modified scaling and squaring method to compute the action of the matrix exponential on the state~\cite{almohy2001computing,jones2001open}.
Convergence of the simulations was verified by using several different time steps, $\Delta t \= 10^{-2} \cdot 2^{-n}$, $n \in \{0, 1, 2, 3, 4\}$.
For a detailed description of a similar numerical setup see Ref.~\onlinecite{innerberger2020electron}.
The density of states was calculated with $H(0)$ from the Fourier transform of 
$-i \theta(t) \left(\expval{c_{i, \uparrow}(t) \, c^\dagger_{i, \uparrow}}{GS} + \expval{c^\dagger_{i, \uparrow} \, c_{i, \uparrow}(t)}{GS}\right)$
(Ref.~\onlinecite{ganahl2015efficient}),
averaging over all sites $i$.

%================================================
\subsection{Quasiparticle picture}
%================================================

\begin{figure} %====================================================
    \centering
    \includegraphics[width=8.6cm]{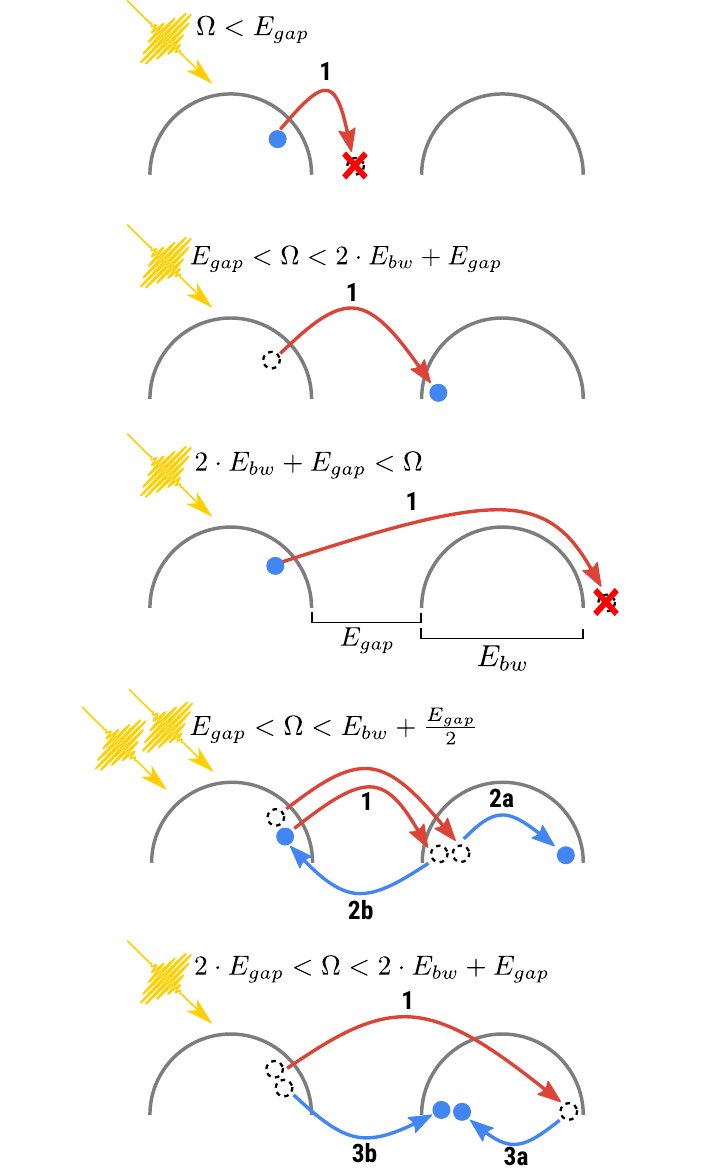}
    \caption{
        Expected excitation processes in the quasiparticle picture, for different ranges of the photon energy $\Omega$. In each subfigure, the horizontal axis represents energy and the semicircles schematically depict the lower and upper Hubbard bands.
        An electron is excited by a photon (step 1) into the upper Hubbard band, given that the photon energy is not too low (first subfigure) or too high (third subfigure).
        For the energy range depicted in the fourth subfigure, Auger recombination can reduce the number of excited electrons (step 2).
        At higher photon energies (last subfigure), 
        a single photon can lead to two excited electrons via impact ionization (step 3).
        \label{fig:impact_ionization} 
    }
\end{figure} %====================================================

The density of states is shown in \cref{fig:hubbard_4x3_dos}.
There is a gap of $E_{gap} \approx 4.7$
and the width of the Hubbard bands is $E_{bw} \approx 5.8$.
The reaction of the system to the pulse will depend on the value of $\Omega$.
The excitations expected to happen in the single particle picture are illustrated in \cref{fig:impact_ionization}.
When $\Omega < E_{gap}$, then the system should not react to the pulse, 
because an incoming photon does not carry enough energy to excite electrons.
When $E_{gap} < \Omega < E_{gap}+2E_{bw}$, a photon can excite an electron
to the upper Hubbard band, leaving behind a hole, %in the lower Hubbard band,
and for $\Omega > E_{gap}+2E_{bw}$, no final state is available, so the photon should not be absorbed.

Electron scattering can later modify the number of excited electrons.
When  $E_{gap} < \Omega < E_{bw} + \frac{E_{gap}}{2}$
(third subfigure of  \cref{fig:impact_ionization}),
Auger recombination~\cite{manousakis2010photovoltaic},
can reduce the number of excited electrons.
Conversely, when $2~E_{gap} < \Omega <  2~E_{bw} + E_{gap}$
(and also $E_{gap} < E_{bw}$),
then an excited electron can subsequently transfer enough energy by scattering processes to excite another electron into the upper Hubbard band. 
Thus a single photon can produce two excited electrons in this process of 
 impact ionization~\cite{manousakis2010photovoltaic,werner2014role,sorantin2018impact,petocchi2019hund,kauch2020enhancement}.

Note that this is an idealized view. In non-equilibrium, the spectral function is time dependent~\cite{eckstein2013photoinduced,hashimoto2016photoinduced,shinjo2017photoinduced,wais2018quantum,okamoto2019timedependent,ishihara2019photoinduced,albota1998design,
kauch2020enhancement,innerberger2020electron,shao2020analysis}
and there is a photo-induced insulator-metal transition in the Hubbard model in various setups~\cite{maeshima2005photoinduced,takahashi2008photoinduced,okamoto2019timedependent,innerberger2020electron,hashimoto2016photoinduced,ishihara2019photoinduced,innerberger2020electron,kauch2020enhancement,lu2015photoinduced,yonemitsu2008theory}.
Such an insulator-metal transition has also been observed in experiment in quasi one- and two-dimensional materials~\cite{iwai2003ultrafast,okamoto2007photoinduced,perfetti2008femtosecond,okamoto2010ultrafast,matsuzaki2015ultrafast,matsuzaki2015photoinduced}.
We also note that the incoming photon pulse has a finite width 
$\tilde\sigma = \frac{1}{2}$ in frequency, 
due to the finite width of the Gaussian peak in the time domain
(see \cref{fig:electric_field}).

%\section{Results}

\begin{figure*} %====================================================
    \centering
    \includegraphics[width=17.2cm]{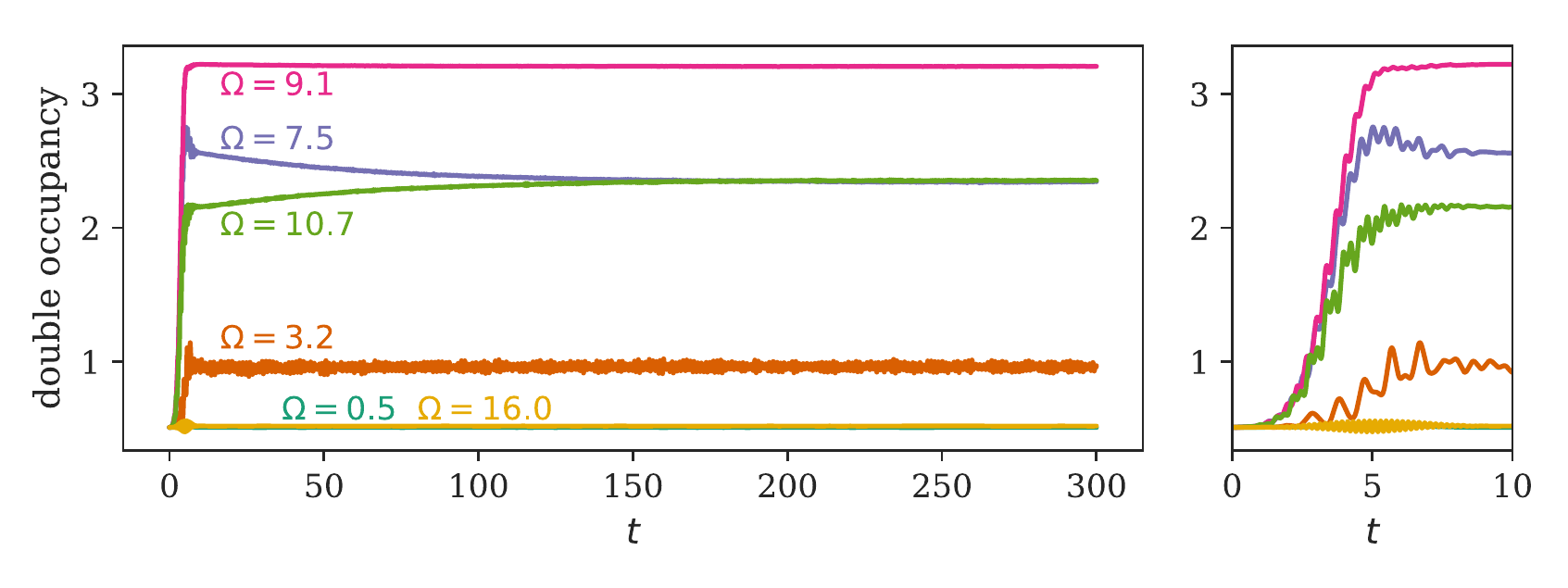}
    \caption{
        Double occupancy as a function of time for $I_0 = 0.71$ and several different $\Omega$.
        The right hand side shows the  double occupancy for short times, where the electric field pulse acts.
        \label{fig:time_series_whole}
    }
\end{figure*} %====================================================

\section{Results}

\begin{figure} %====================================================
    \centering
    \includegraphics[width=8.6cm]{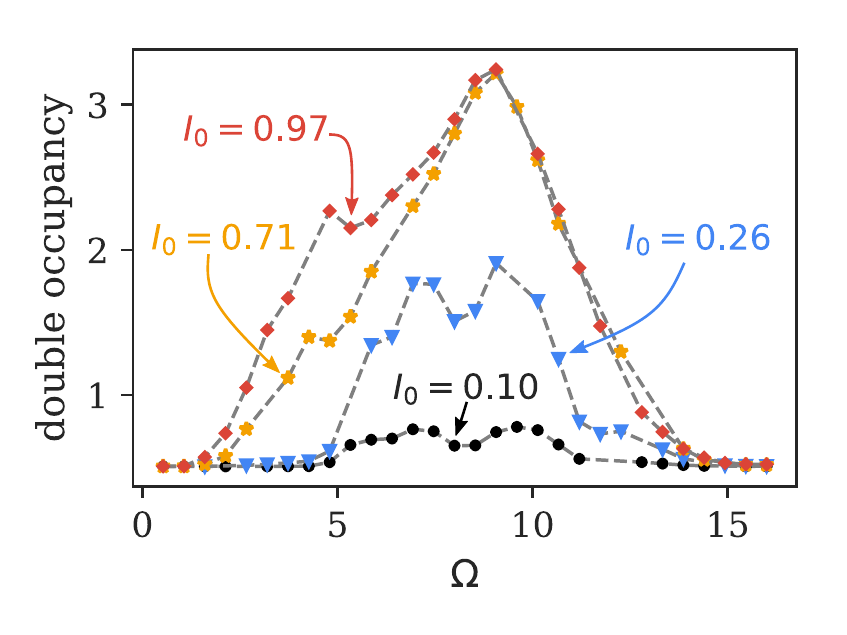}
    \caption{
        Double occupancy at $t=20$ after the pulse as a function of $\Omega$ and several different intensities $I_0$.
        For small $I_0$ the response is in good agreement with the quasiparticle picture according to the density of states.
        To minimize small oscillations with respect to time we show averages over times from $t=17.5$ to $t=22.5$.
        Color codes for the intensity $I_0$.
        \label{fig:double_occ_inital_response}
    }
\end{figure} %====================================================

\begin{figure} %====================================================
    \centering
    \includegraphics[width=8.6cm]{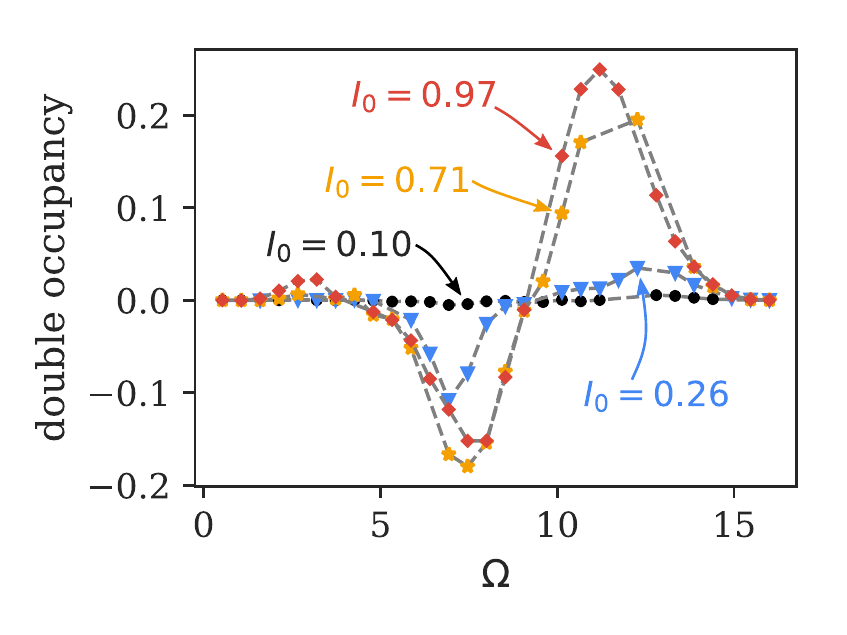}
    \caption{
        Further change of double occupancy after the pulse.
        Shown is the difference between the values at $t = 300$ and $t = 20$.
        To minimize small oscillations with respect to time we 
        used occupancies averaged over intervals with length $5$ around $t=20$ resp.\ $t=297.5$.
        Color codes for the intensity $I_0$.
        \label{fig:double_occ_change}
    }
\end{figure} %====================================================

%=================================================
\subsection{Time evolution of the double occupancy} \label{sec:time_evolution_double_occupancy}
%=================================================

In \cref{fig:time_series_whole} we show the time evolution of the double occupancy for several photon energies $\Omega$
at an intensity of $I_0=0.71$.
The overall behavior is mostly in agreement with expectations from the quasiparticle picture,
except for $\Omega=3.2$ below the gap.

For all $\Omega$, there are initial oscillations with a frequency of roughly 
$2\Omega$. At large times, the double occupancy converges to a constant value, up to small fluctuations.

For very small and very large photon energies, the double occupancy oscillates slightly during the pulse
but then returns to almost exactly its original value. 

At frequencies $\Omega$ between $E_{gap}$ and $E_{gap}+2E_{bw}$, 
where absorption is expected,
the double occupancy rises quickly during the pulse, 
i.e.,  elecrons are excited across the gap and energy is transferred into the system. 
Note that after the pulse has decayed, the Hamiltonian $H(t)$ becomes the time-independent bare $H(0)$ again, so that energy is then conserved.
At $\Omega=9.1$, just below the region where impact ionization is expected, the double occupancy stays (almost) constant after the pulse.
At $\Omega=10.7$, the double occupancy shows the expected impact ionization behavior, noticeably rising further after the pulse,
which was also observed by Kauch et al.\cite{kauch2020enhancement}.
We note that the corresponding time scale is large compared to the electron hopping time~\cite{werner2014role,petocchi2019hund,kauch2020enhancement}

At $\Omega=7.5$, 
the double occupancy goes down after the pulse,
compatible with the expected  Auger recombination, 
beginning already after the pulse maximum at time $t=5$.

Remarkably, at $\Omega=3.2$, below $E_{gap}$, where the quasiparticle picture would forbid excitations, \cref{fig:time_series_whole} exhibits a sizeable increase of the double occupancy. We will later show that this is due to multiphoton excitations.

In \cref{fig:double_occ_inital_response} we display the double occupancy 
at $t=20$ after the pulse, for different intensities $I_0$  and different frequencies $\Omega$.
There is a strong nonlinear dependence on $I_0$. We will show 
in \cref{sec:eigenstate_spectrum} 
that it occurs together with  multiple photon excitations.
Indeed, for $\Omega$ below $E_{gap}$, a sizeable excitation of double occupations does not occur for small intensities, but only at large $I_0$, as would be expected for
multiphoton absorption.

The energy absorbed by the Hubbard system matches the change in double occupancy in \cref{fig:double_occ_inital_response} closely,
which will be further explored in \cref{sec:eigenstate_thermalization}.

In \cref{fig:double_occ_change} we show the further change of double occupancy after the pulse, from $t=20$ to $t=300$ where it has converged well for all $\Omega$.
In the energy range $2~E_{gap} < \Omega < E_{gap} + 2~E_{bw}$
the double occupancy increases, i.e., there is impact ionization, 
as suggested by the quasiparticle picture.
Impact ionization is larger at higher intensities. We will show below that this is again connected to the absorption of several photons during the pulse.
Notably, when the intensity is large, impact ionization even occurs at small  $\Omega < E_{gap}$.
The double occupancy decreases after the pulse in a range of lower values of $\Omega$. For the small $I_0$, this range closely matches the expectation 
 $E_{gap} < \Omega < E_{bw} + \frac{E_{gap}}{2}$ from the single particle picture, while at $I_0=0.71$ and $I_0=0.97$, the range extends to larger energies.
We will discuss these processes further in \cref{sec:TimeEvolIndividPeaks}.

%=================================================
\subsection{Eigenstate spectrum} \label{sec:eigenstate_spectrum}
%=================================================

\begin{figure} %====================================================
    \centering
    \includegraphics[width=8.6cm]{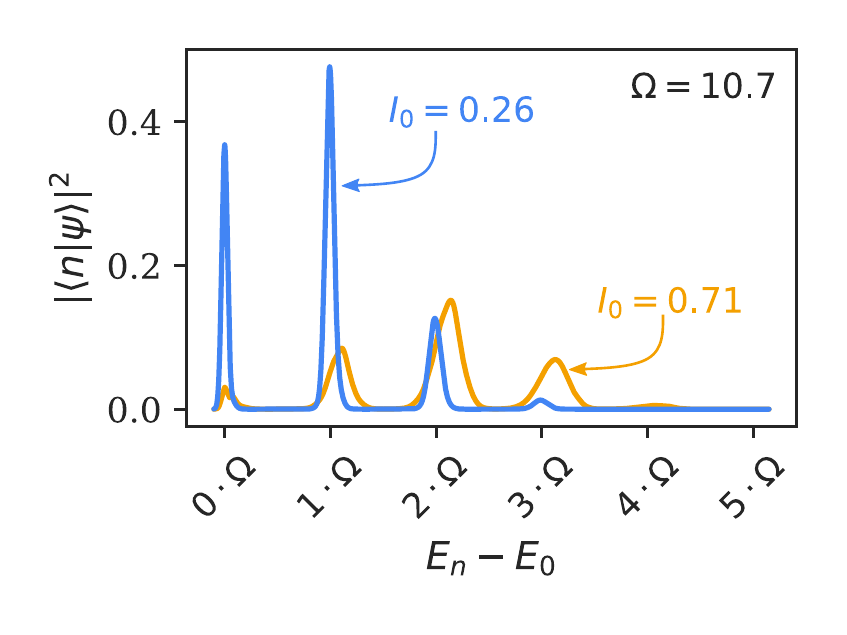}
    \caption{
        Eigenstate spectrum at $t = 20$ for $\Omega=10.7> 2  E_{gap}$ at
        two values of the electric field intensity.
        \label{fig:loschmidt_large_omega}
    }
\end{figure} %====================================================

\begin{figure} %====================================================
    \centering
    \includegraphics[width=8.6cm]{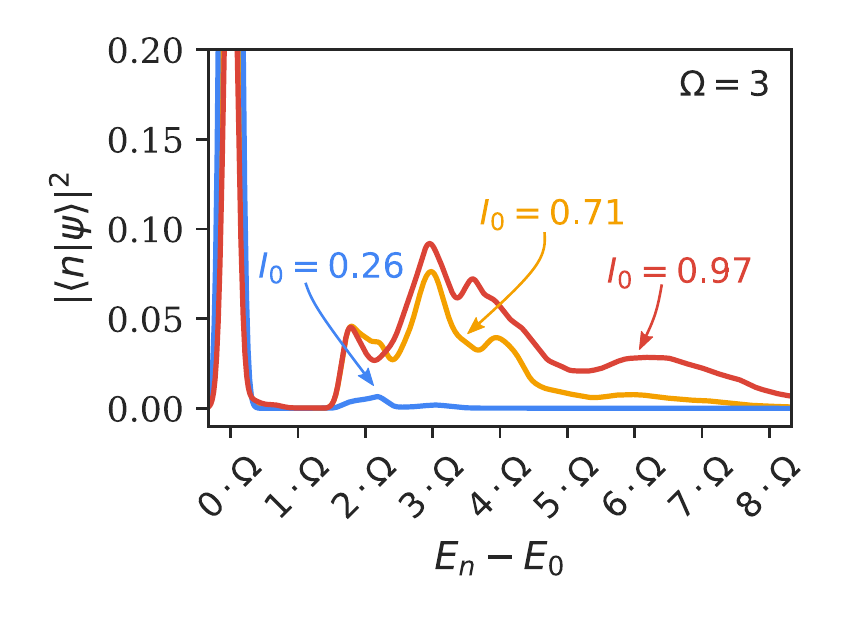}
    \caption{
        Eigenstate spectrum at $t = 20$ for  $\Omega=3<E_{gap}$ at different
        field intensities.
        %three values of the electric field intensity at
        The first peak at $0\cdot\Omega$ is dominant for all three cases.
        Its height is $\approx 1.4$ for $I_0 = 0.26$,  $\approx 0.8$ for $I_0 = 0.71$, and $\approx 0.3$ for $I_0 = {0.97}$.
        \label{fig:loschmidt_small_omega}
    }
\end{figure} %====================================================

For a better understanding of the excited state after the pulse, we examine the eigenstate spectrum of $\ket{\psi(t)}$.
To this end we compute the Fourier transform of the Loschmidt amplitude, with respect to an auxiliary time span $\tau$ (Refs.~\onlinecite{rylands2019loschmidt,kennes2020loschmidt,maislinger2018conference,LoschmidtIndependent}).
\begin{subequations}
    \begin{align}
      L(\tau) &= \bra{\psi(t)} ~ e^{-i \tau H(0)} ~ \ket{\psi(t)} \label{eq:loschmidt_time_domain} \\
    \int_{-\infty}^{\infty} e^{i \omega \tau} L(\tau) \dd{\tau} 
     &= 2 \pi ~ \sum_n \left| \braket{n}{\psi(t)} \right|^2 ~ \delta(\omega - E_n) \label{eq:fourier_transform_loschmidt}
    \end{align}
\end{subequations}
 \Cref{eq:fourier_transform_loschmidt} can also be viewed as the probability distribution of work done on the system~\cite{silva2008statistics}
 by the electric field.
In \cref{eq:fourier_transform_loschmidt}, 
the sum runs over all eigenstates  $\ket{n}$ of $H(0)$, and $E_n$ is the energy of $\ket{n}$.
Through \cref{eq:fourier_transform_loschmidt} we obtain the energy spectrum of an arbitrary state, while only using the tool of time-evolution.
This is very useful here, because the large Hilbert space dimension prohibits full diagonalization of the Hamiltonian.
We note that the eigenstate spectrum is independent of time after the pulse has decayed, because then the Hamiltonian $H(t)$ reverts back to the initial Hamiltonian $H(0)$.

We dampened the time evolution in \cref{eq:loschmidt_time_domain} with $\exp(-t^2/(2\sigma^2))$, $\sigma=5/\sqrt{2}$, thus widening the spectra by $\tilde{\sigma}=1/\sigma\approx 0.3$.

In \cref{fig:loschmidt_large_omega} we show the resulting spectrum for the case of $\Omega=10.7$, after the pulse has decayed.
The spectrum has a very distinct peaked structure, with the distances between the peaks close to $\Omega$.
The excited state $\ket{\psi(t)}$ thus consists of groups of eigenstates of $H(0)$ close to multiples of the photon energy.
At low intensity, $I_0=0.26$, the peaks are narrow, and excitations at $1\Omega$ and $2\Omega$ dominate.
For large intensity $I_0=0.71$, the peaks are wider, shifted slightly 
to higher energies, and they include higher multiples of $\Omega$.
The ground state is then almost depleted, in line with the saturation of
double occupancy in \cref{fig:double_occ_inital_response}.
Since $2\Omega$ is larger than $E_{gap}+2E_{bw}$ here,
peaks beyond $1\Omega$ 
must correspond to the sequential excitation of several electrons from the lower to the upper Hubbard band, 
by several photons.
We will investigate these peaks individually in \cref{sec:TimeEvolIndividPeaks}.
%
%% OMEGA=3:
Below the gap, $\Omega<E_{gap}$, 
\cref{fig:double_occ_inital_response} showed that
the double occupancy is increased by the light pulse
when the intensity $I_0$ is large, even though single photon absorption is energetically forbidden.
In \cref{fig:loschmidt_small_omega} we show corresponding eigenstate spectra for $\Omega=3$ at different intensities $I_0$.
At small $I_0=0.26$, where there is very little absorption, \cref{fig:loschmidt_small_omega} shows that
$\ket{\psi(t)}$ has almost returned to the ground state (slightly widened in the plot), with an additional small excitation at $2\Omega$.
At larger intensity $I_0=0.71$, however, states with higher multiples of the photon energy have become excited. Thus the increase in double occupancy here is indeed due to multiphoton absorption.
The largest amplitude in  \cref{fig:loschmidt_small_omega} is from three-photon excitations at $3\Omega = 9$,
where the available phase space is the largest as indicated by the highest
absorption values in \cref{fig:time_series_whole,fig:double_occ_inital_response}.
At excitation energy $4\Omega$, both 4-photon absorption and two sequential two-photon absorptions can contribute.
At $5\Omega=15$ and beyond, the end of the bandwidth has been reached, so that
the excitations, still sizeable in  \cref{fig:loschmidt_small_omega},
must correspond to sequential absorptions.
We note that since the density of states develops a small in-gap density after electron excitations ~\cite{eckstein2013photoinduced,hashimoto2016photoinduced,okamoto2019timedependent,kauch2020enhancement,innerberger2020electron},
additional processes with single-photon absorption will also be possible
after the initial two-photon absorption has taken place.

%==================================================================
\subsection{Time evolution of individual photon-absorption peaks}\label{sec:TimeEvolIndividPeaks}
%==================================================================

\begin{figure*} %====================================================
    \centering
    \includegraphics[width=17.2cm]{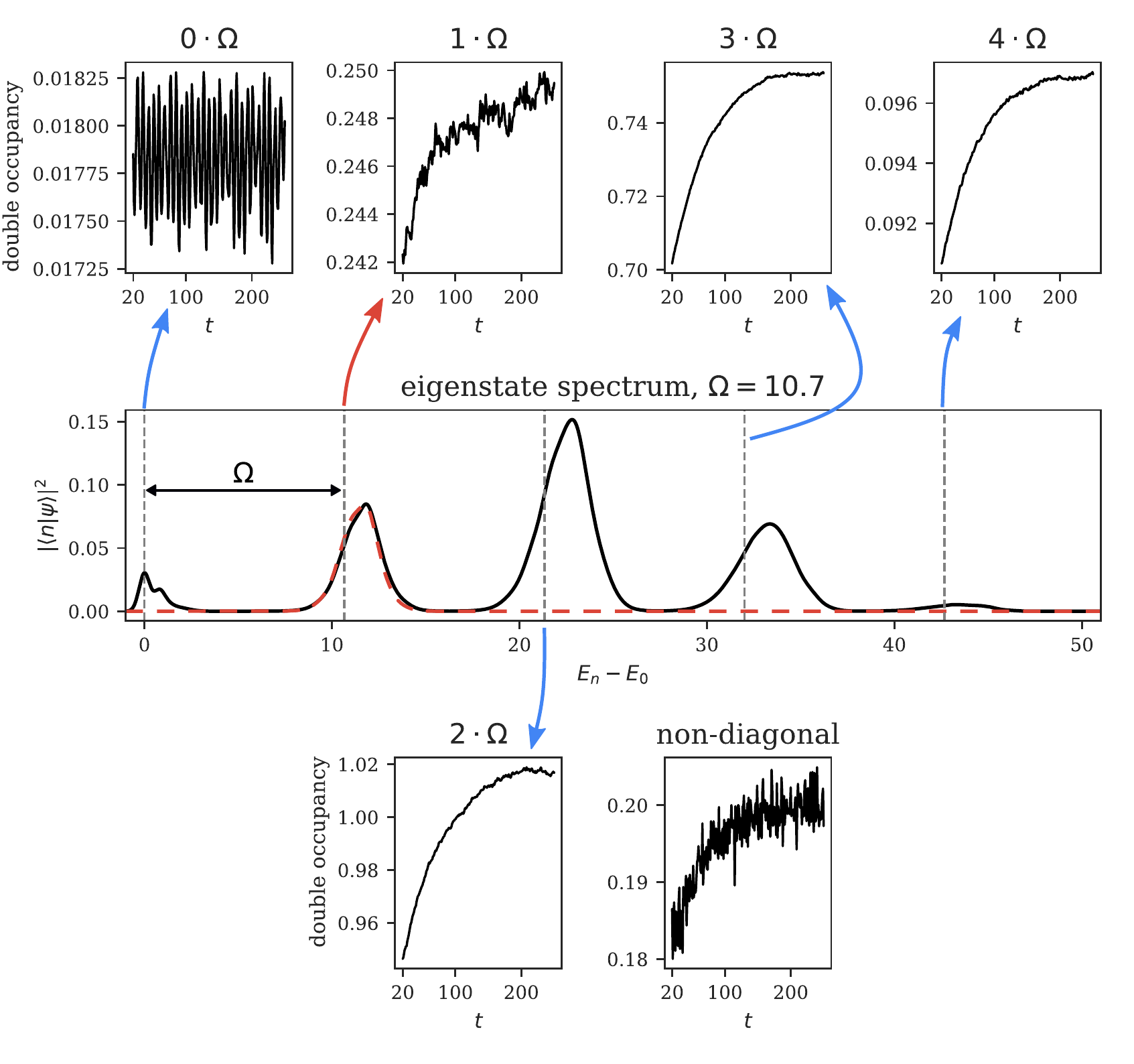}
    \caption{
        Double occupancy of the different peaks as a function of time, 
        computed with \cref{eq:filtered_expectation_value,eq:filtered_expectation_value_nondiag}, for $\Omega = 10.7$ and $I_0 = 0.71$.
        The center row displays the eigenstate spectrum computed with \cref{eq:fourier_transform_loschmidt}, showing peaks with a distance of $\approx \Omega$.
        The red dashed curve in the center row shows the eigenstate spectrum after filtering for one specific peak.
        The top and bottom rows display the contributions of the filtered peaks to the double occupancy.
        \label{fig:double_occ_peaks_11}
    }
\end{figure*} %====================================================

\begin{figure*} %====================================================
    \centering
    \includegraphics[width=17.2cm]{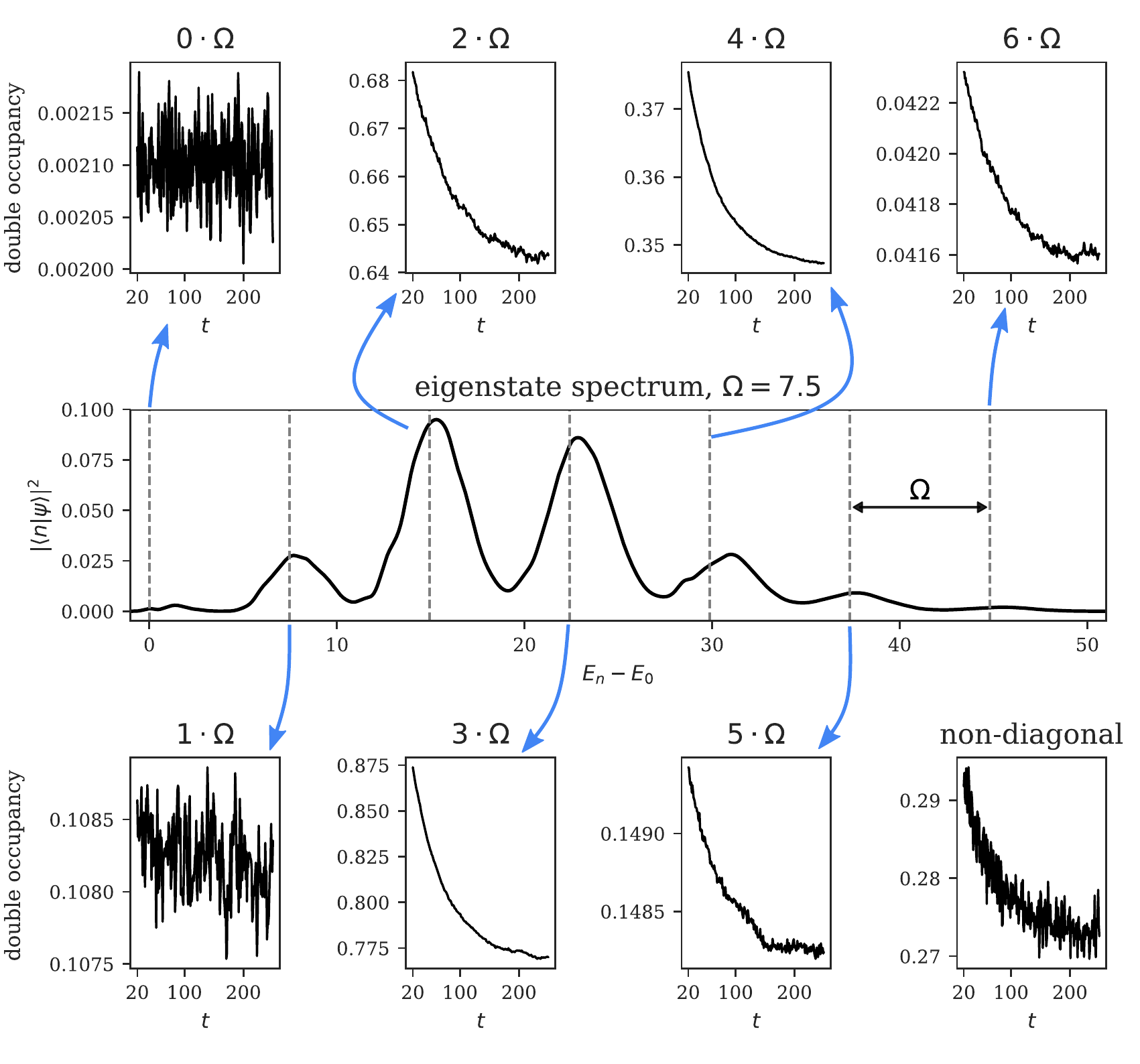}
    \caption{
        Double occupancy of the different peaks as a function of time, 
        computed with \cref{eq:filtered_expectation_value,eq:filtered_expectation_value_nondiag}, for $\Omega = 7.5$ and $I_0 = 0.71$.
        The center row displays the eigenstate spectrum computed with \cref{eq:fourier_transform_loschmidt}, showing peaks with a distance of $\approx \Omega$.
        The top and bottom rows display the contributions of the filtered peaks to the double occupancy.
        \label{fig:double_occ_peaks_07}
    }
\end{figure*} %====================================================

Knowing the structure of the eigenstate spectrum, one can gain considerable additional insight into the development of $\ket{\psi(t)}$.
Here we introduce a technique to 
decompose a state $\ket{\psi(t)}$ into states $\ket{f_i(t)}$ with support around the individual photon peaks.
We isolate the contribution of a photon peak to $\ket{\psi(t)}$ by applying a filter,

\begin{equation}
  \ket{\varphi(t, \sigma_f, E_f)} ~=~ \frac{1}{\mathcal{N}(t)} ~ e^{-\frac{(H(0) - E_f)^2}{2 \sigma_f^2}} ~ \ket{\psi(t)} ~.
  \label{eq:filter}
\end{equation}

We take $\mathcal{N}(t)$ such that $\braket{\varphi(t)}{\varphi(t)} = 1$.
This is a Gaussian peak centered around $E_f$.
Expanding $\ket{\varphi(t)}$ in the eigenbasis of $H(0)$ shows that basis states which are energetically too far away from $E_f$ are discarded  in $\ket{\varphi(t)}$.
Note that the filtering  from $\ket{\psi(t)}$ to $\ket{\varphi(t)}$ is very similar to doing a time-evolution.

Suppose that we have derived  a number of distinct states $\ket{f_i(t)}$
from a state $\ket{\psi(t)}$  with the filtering procedure above.
To find the best representation $\ket{\psi'(t)}$ of $\ket{\psi(t)}$ that is a linear combination of filter results $\ket{f_i(t)}$ we use:

\begin{equation}
    \ket{\psi'(t)} = \sum_i \alpha_i(t) \ket{f_i(t)} 
\end{equation}

and choose the coefficients $\alpha_i(t)$ such that $\big| \ket{\psi(t)} - \ket{\psi'(t)} \big|^2$ is minimized.
We will look at the expectation value of an operator $O$ with respect to the single states $\ket{f_i(t)}$,
including the coefficients $\alpha_i(t)$:

\begin{subequations}
    \begin{align}
        \expval{O}_i &= \left| \alpha_i(t) \right|^2 \expval{O}{f_i(t)} \label{eq:filtered_expectation_value} \\
        \expval{O}_{\text{non-diag}} &= \sum_{i \neq j} \bar{\alpha}_i(t) \, \alpha_j(t) \, \matrixel{f_i(t)}{O}{f_j(t)} \label{eq:filtered_expectation_value_nondiag}
    \end{align}
\end{subequations}

(with the time dependence not denoted). Specifically for the photon peaks we define $\ket{f_i(t)}$ as follows:

\begin{equation}
    \ket{f_i(t)} = \ket{\varphi\left(t, \frac{\Omega}{3}, E_{GS} + i \cdot \Omega\right)}
    \label{eq:used_filter_definition}
\end{equation}

In other words, $\ket{f_i(t)}$ is a filtered state centered around the energy that is $i$ times $\Omega$ above the ground state energy, filtered with a width of $\sigma_f = \frac{\Omega}{3}$.
Note that each filtered state still contains many eigenstates,
so that observables, like the double occupancy, remain time dependent even after the light pulse.

%For the cases investigated, shown below,
We find that the set of filtered states $ \ket{f_i(t)}$ 
provides a good representation of the original state $\ket{\psi(t)}$ at all times.
The absolute value of the overlap between the filtered states $\left|\braket{f_i(t)}{f_j(t)}\right|$ is about $5 \cdot 10^{-2}$ when they are $  \Omega$ apart, $ 10^{-4}$ when $2  \Omega$ apart, and $ 10^{-8}$ when they are $3  \Omega$ apart.
The norm squared of the difference between the original time-evolved state and the best approximation 
$\big|\ket{\psi(t)} - \sum_i \alpha_i(t) \ket{f_i(t)}\big|^2$ 
is of the order of $10^{-2}$.
The relative difference between the expectation value of the double occupancy of the original states and the sum of filtered state contributions \cref{eq:filtered_expectation_value,eq:filtered_expectation_value_nondiag} is of the order of $0.5~\%$.

We examine two cases, $\Omega=10.7$ where the double occupancy increases, and $\Omega=7.5$ where it decreases after the pulse.
In both cases we find that during the pulse, the initial rise of the photon peaks occurs sequentially, delayed by roughly one hopping time for each additional photon, 
in line with the expected sequential nature of photon absorptions at these values of $\Omega$.

The eigenstate spectra of the two states after the pulse are shown in \cref{fig:double_occ_peaks_07,fig:double_occ_peaks_11} (center row).
At every time $t$, we separately filter out each peak and calculate the double occupancy according to \cref{eq:filtered_expectation_value},
including the coefficient $|\alpha_i(t)|^2$.
The results are shown in \cref{fig:double_occ_peaks_11} and \cref{fig:double_occ_peaks_07} (top and bottom rows).
%

% Omega=10.7 %%%%%%%%%%%%%%%%%%%%%
For $\Omega = 10.7$,  we saw in \cref{fig:time_series_whole} that there is an increase of double occupancy after the pulse,
in line with the impact ionization expected in the quasiparticle picture.
\Cref{fig:double_occ_peaks_11} shows that indeed each photon peak 
and also the non-diagonal part contribute to the increase.
The largest contributions to the double occupancy
and to its increase come from the states with two and three absorbed photons,
each of which can separately contribute to the impact ionization process sketched in \cref{fig:impact_ionization}. 

% Omega=7.5 %%%%%%%%%%%%%%%%%%%%%
At $\Omega = 7.5$,  absorption of up to 6 photons is visible, and the remaining contribution of the ground state is tiny.
Now $\ket{\psi(t)}$  has a decreasing double occupancy as a function of time.
The individual peaks in \cref{fig:double_occ_peaks_07} all contribute to this decrease,
with the notable exception of the single photon peak, for which the double occupancy stays almost constant.

This difference in behavior is in line with  Auger recombination %~\cite{manousakis2010photovoltaic} 
in the quasiparticle picture, shown in \cref{fig:impact_ionization} (fourth subfigure), 
which is only possible when at least two electrons are excited into the upper Hubbard band. Thus this decay channel is absent in the single photon peak.
The upper bound for this process is 
$\Omega<E_{bw} + \frac{E_{gap}}{2} \approx 8.1$ for two initially excited electrons,
matching the observed $\Omega$-range of decay at small intensities in \cref{fig:double_occ_change}.
The range in \cref{fig:double_occ_change}
is wider at larger intensities, $I_0=0.71$ and $0.97$, where absorption of more photons becomes important (\cref{fig:double_occ_peaks_07,fig:loschmidt_large_omega}),
thus allowing Auger-like processes with more initial photons and larger energy range.
Furthermore, for more initially excited electrons there are more decay channels, suggesting a stronger decrease of double occupancy, in agreement with the behavior of the large $3\Omega$ peak in \cref{fig:double_occ_peaks_07}, which shows the largest change in double occupancy after the pulse.

%=================================================
\subsection{Eigenstate Thermalization} \label{sec:eigenstate_thermalization}
%=================================================

\begin{figure*} %====================================================
    \centering
     \includegraphics[width=17.2cm]{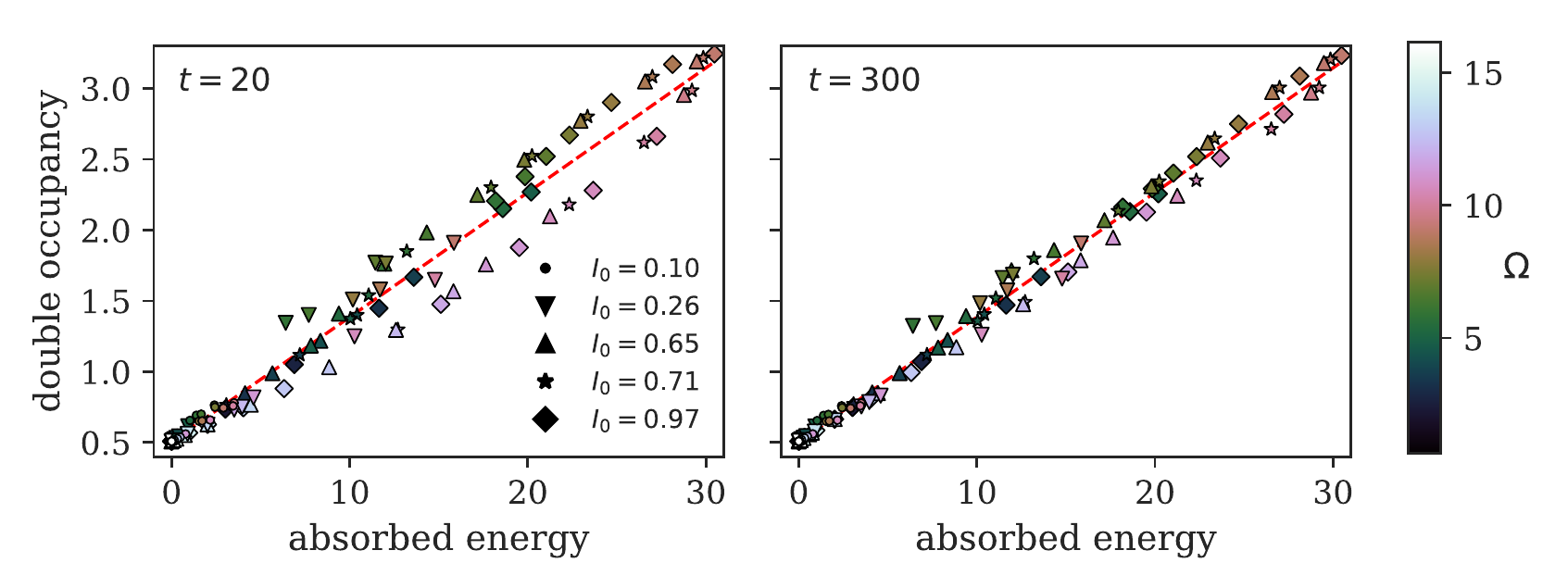}
    \caption{
        Double occupancy as a function of absorbed energy from the pulse, for time $t=20$ after the pulse and for $t=300$ after convergence.
        Different colors mark different pulse frequencies $\Omega$.
        For long simulation times, the double occupancies tend to a linear function.
        To reduce the effect of short-time oscillations, the figures show averages over the double occupancy in  time intervals of length $5$ around $t=20$, resp. $t=297.5$.
        }
\label{fig:energy_double_occ_both}
    
\end{figure*} %====================================================

\begin{figure} %====================================================
    \centering
    \includegraphics[width=8.6cm]{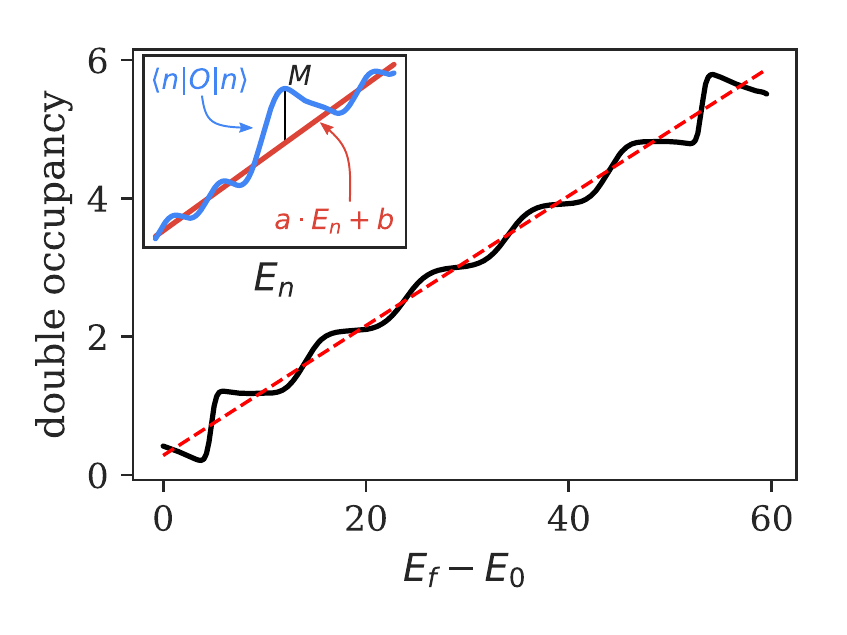}
    \caption{
        Double occupancy as a function of energy for uniformly sampled states of the $4\times 3$ Hubbard model, each projected onto a narrow energy range around energies $E_f$.
        The slope of the linear fit agrees to within a few percent
        with the one in \cref{fig:energy_double_occ_both}.
       Inset: Sketch of expectation values of an operator $O$ that is well approximated by a linear function with respect to the eigenenergies $E_n$.
        \label{fig:eth_expectation_value}
    }
\end{figure} %====================================================

Another way to learn about the photoexcited system is to look at the double occupancy as a function of absorbed energy, shown in \cref{fig:energy_double_occ_both}.
Interestingly, whereas the double occupancy 
at time $t=20$ depends on intensity and frequency separately,
after convergence at long simulation times it depends almost only on the absorbed energy,
with an almost linear relation between those two quantities.

 In many physical systems it has been observed that long time averages of some expectation values are actually close to a  
microcanonical average, which has been discussed under the name of Eigenstate Thermalization Hypothesis (ETH)~\cite{deutsch1991quantum,srednicki1994chaos,deutsch2018eigenstate}.

When the Hamiltonian is not time-dependent (long after the light pulse in our case), a quantum state under unitary time-evolution can be written in the eigenbasis of the Hamiltonian:

\begin{equation}
    \ket{\psi(t)} = \sum_n e^{-i (t-t_0) E_n} \, \alpha_n \, \ket{n}, ~~~ \alpha_n = \braket{n}{\psi(t_0)}
\end{equation}

The mean of the expectation value of an observable $O$ is

\begin{subequations}
    \begin{align}
        \overline{\expval{O(t)}} &:= \frac{1}{t - t_0} \int_{t_0}^t \, \expval{O(t')} \, \dd{t'} \\
        &= \sum_{m, n} \alpha_m^* \alpha_n \matrixel{m}{O}{n} \, \frac{\int_{t_0}^t \, e^{-i (t'-t_0) (E_n - E_m)} \, \dd{t'}}{t-t_0} 
    \end{align}
\end{subequations}

Under the assumption of no degeneracy, %one can easily verify that 
only the diagonal contributions survive for long times:

\begin{equation}
    \overline{\expval{O(t)}} \xrightarrow[t \, \gg \, t_0]{} \sum_n \expval{O}{n} \, \left|\braket{n}{\psi}\right|^2
    \label{eq:eth_long_time_expectation_value}
\end{equation}

The Eigenstate Thermalization Hypothesis \cite{deutsch1991quantum,srednicki1994chaos,deutsch2018eigenstate} tries to explain the peculiar observation that this long-time mean of $\expval{O(t)}$ often coincides with a microcanonical average, where we take the average of the expectation values of the eigenstates in a small energy window $\Delta E$ around
$E_\psi=\sum_n E_n  \left|\braket{n}{\psi}\right|^2$,
For $t\gg t_0$:

\begin{align}
     %t \gg t_0: ~ 
     \overline{\expval{O(t)}} ~\stackrel{?}{=}~ 
     \expval{O}_{mc} \equiv \frac{1}{\mathcal{N}} \sum_{\substack{n \\ \left|E_n - E_{\psi}\right| < \Delta E}} \expval{O}{n}
     \label{eq:ETHmc}
\end{align}

This is rather surprising, because the left hand side of the equation above depends on the specific coefficients of the state in the eigenbasis $\left|\braket{n}{\psi}\right|^2$, while the right hand side does not.
One possible explanation 
is that for many physical situations the state has a single peak in the eigenstate spectrum and 
in addition, the difference between two expectation values of two separate eigenstates is small when the difference between their eigenenergies is small:

\begin{equation}
    E_m \approx E_n ~ \Rightarrow ~ \expval{O}{m} \approx \expval{O}{n}
\end{equation}

In our case, however, we do not have a single peak in the eigenstate spectrum, but a series of peaks that stretch over almost the whole eigenenergy range, as can be seen in \cref{fig:double_occ_peaks_07,fig:double_occ_peaks_11}.
We now show that eigenstate thermalization still holds when there is an almost linear relation between the eigenenergies and the expectation values of their respective eigenvalues (\cref{fig:eth_expectation_value}, inset) in the energy range(s) where $\ket{\psi(t_0)}$ has an overlap with the eigenstates:

\begin{subequations}
    \begin{align}
        \expval{O}{n} &= a \cdot E_n + b + \epsilon(E_n)  \label{eq:OnLinearInEn} \\
        \left| \epsilon(E_n) \right| &\leq M
        \label{eq:OnLinearInEnDeviation}
    \end{align}
\end{subequations}

with $M$ the maximum deviation from the linear behavior.
Then the long-time average of $\expval{O(t)}$ 
has the same linear behavior 
as a function of $E_\psi$ %\sum_n E_n  \left|\braket{n}{\psi}\right|^2$,
with at most the same maximum deviation $M$, no matter how peaked the structure of $\ket{\psi(t_0)}$ is in the eigenstate spectrum.
Namely, for large $t\gg t_0$, where \cref{eq:eth_long_time_expectation_value} holds, we have

\begin{align}
    &\left|\overline{\expval{O(t)}} - a \, E_\psi - b\right| = \nonumber \\
    &= \left| \sum_n \expval{O}{n} \, \left|\braket{n}{\psi}\right|^2 - a \, E_\psi - b \right| \nonumber \\
    &= \left| \sum_n \left(a \cdot E_n + b + \epsilon(E_n)\right) \, \left|\braket{n}{\psi}\right|^2 - a \, E_\psi - b \right|\nonumber \\
    &= \left| \sum_n \epsilon(E_n) \left|\braket{n}{\psi}\right|^2 \right| 
    ~\leq~ M
    \label{eq:ETHlinEpsi}
\end{align}

Similarly, for the difference between the long time average and the microcanonical average \cref{eq:ETHmc}

\begin{equation}
    \left|\overline{\expval{O(t)}} - \expval{O}_{mc} \right|
    ~\leq~ M + M'+ |a|\frac{\Delta E}{\cal N}
\end{equation}

with $\Delta E$ and $\cal N$ from \cref{eq:ETHmc}
and $M'$ the maximum deviation from linear behaviour within the small energy range $\Delta E$.
The upper bounds apply when $\ket{\psi}$ has support only from eigenstates where the deviation \cref{eq:OnLinearInEnDeviation} is maximal.

We examined the linearity \cref{eq:OnLinearInEn} of the double occupancy
in the Hubbard model
by uniformly sampling states from the hypersphere of normed states~\footnote{We used Marsaglia's method~\cite{marsaglia1972choosing}, sampling in the occupation number basis.
Because the transformation to the eigenbasis is unitary, the states are also drawn uniformly from the hypersphere of normed states in the eigenbasis.}.
For each value of $E_f$ in \cref{fig:eth_expectation_value}, we took a sample of $N_r=9$ states $\ket{r}$, projected and normalized each state to a small range around $E_f$ by
$\ket{r_f}=\frac{1}{\cal{N}}e^-\frac{(H-E_f)^2}{2}\ket{r}$, and
plotted the averaged double occupancy 
$d(E_f)= \frac{1}{N_r}\sum_{r_f}\expval{\sum_i n_{i\uparrow}n_i\downarrow}{r_f}$,
which is similar to the microcanonical average \cref{eq:ETHmc}, since contributions that are nondiagonal in eigenstates cancel stochastically.
The approximately linear behavior in \cref{fig:eth_expectation_value} was
verified for a smaller $3 \times 2$ system where exact diagonalization is still possible.

\Cref{fig:eth_expectation_value} indicates that for the Hubbard model,
the double occupancy of eigenstates is indeed close to linear in the eigenstate energies, like \cref{eq:OnLinearInEn},
with the same slope as in \cref{fig:energy_double_occ_both}.
The steps in the figure correspond to individual double occupations.

In our case, the double occupancy after the light pulse converges (up to small fluctuations) at large times $t$. The converged value, shown in \cref{fig:energy_double_occ_both},
is then the same as the long time average.
Taking into account the actual eigenstate spectra of the excited states,
the convergence towards linear behavior  in \cref{fig:energy_double_occ_both}
indeed corresponds to \cref{eq:ETHlinEpsi}.

\section{Summary}

 We investigated the non-equilibrium response of a strongly correlated Mott insulator to a short light pulse, using exact-diagonalization based  calculations on a $4 \times 3$ Hubbard model
 for a large range of light intensities and of photon energies $\Omega$.
 The pulse excites electrons into the upper Hubbard band,
 quickly increasing the number of doubly occupied sites.
 At sufficiently large photon energies, we observed impact ionization 
 (also seen in Ref.~\onlinecite{kauch2020enhancement}), 
 namely a further increase in the double occupancy over time
 after the light pulse had ended.
 Conversely, at lower photon energies, 
 we observed Auger recombination
 with a reduction in double occupancy.

 We calculated the eigenstate spectra of the non-equilibrium states,
 i.e.\ the probability distribution of the work done by the light pulse,
 as the Fourier transform of the Loschmidt amplitude.
 The resulting spectra exhibit distinct peaks at distances of about $\Omega$,
 corresponding to the absorption of multiple photons.
The absorption rate is strongly nonlinear in light intensity.
 Multiphoton absorption was identified
 for small photon energies below the band gap,
 which leads to electron excitations  when the light intensity is large.

We introduced a technique, 
using tools similar to time evolution, 
to isolate photon peaks in the eigenstate spectra. This enabled us to investigate the non-equilibrium evolution of double occupancy in individual photon peaks. We showed for example that, as expected from the quasiparticle picture of Auger recombination, double occupations excited by a single photon do not contribute to the overall reduction over time which occurs at intermediate light frequencies.

We found that at large times, the double occupancy moves towards a function that only depends on the absorbed energy, reminiscent of the Eigenstate Thermalization Hypothesis. 
Eigenstate Thermalization is usually observed for states with support in a narrow region of eigenenergies. We showed that the observed dependence on energy alone will also happen for states which contain a wide range of eigenenergies, when the relevant observable is almost a linear function of energy.

 The analysis of  eigenstate spectra via the Loschmidt amplitude and the filtering of the relevant energy ranges provided valuable insight into multiple photon absorptions and should prove to be useful tools to investigate strongly correlated systems, when full diagonalization is not possible, but the computation of the time-evolution is accessible.

\section*{Acknowledgements}
The authors acknowledge financial support by the Austrian Science Fund (FWF) through SFB ViCoM, F4104.
We thank K. Held, A. Kauch, and D. Bauernfeind for stimulating and fruitful initial discussions. The computational resources were provided by TU Graz.

\bibliography{bibdata}

\begin{thebibliography}{75}
\bibitem [{Los()}]{LoschmidtIndependent}%
  \BibitemOpen
  \href@noop {} {}\bibinfo {note} {The analysis of eigenstate spectra via the
  Loschmidt amplitude was independently developed for the present
  study}\BibitemShut {NoStop}%
\end{thebibliography}

\end{document}